# Polarization Properties of the Cholesteric Liquid Crystals

A.H. Gevorgyan [1], M.Z. Harutyunyan [1], K.B. Oganesyan [2,3*], E.A. Ayryan [3], Michal Hnatic [4,5,6], M.S. Rafayelyan [7], Yuri V. Rostovtsev [8], Gershon Kurizki [9]

[1] Yerevan State University, Yerevan, Armenia
[2] Alikhanian National Lab, Yerevan Physics Institute, Yerevan, Armenia
[3] LIT, Joint Institute for Nuclear Research, Dubna, Russia
[4] Faculty of Sciences, P. J. Safarik University, Park Angelinum 9, 041 54 Kosice, Slovakia
[5] Institute of Experimental Physics SAS, Watsonova 47, 040 01 Kosice, Slovakia ˇ
[6] BLTP, Joint Institute for Nuclear Research, Dubna, Russia
[7] Université Bordeaux 1 , Talence, France
[8] University of North Texas, Denton, TX, USA
[9] Weizmann Institute of Science, Rehovot, Israel

[*] bsk@yerphi.am

In the present paper, we investigate the polarization properties of the cholesteric liquid crystals (CLCs) with an isotropic/anisotropic defect inside them. Possibilities of amplification of the polarization plane rotation and stabilization of the light polarization azimuth by these systems are investigated in details.

## I. INTRODUCTION

In recent years, considerable interest has been attracted to the photonic crystals (PCs) [1-4], which are a special class of artificial and self organizing structures with periodic changes of spatial dielectric properties in the scale of the optical order of wavelength. Such media are also called *photonic band-gap* (PBG) systems, since there is a zone of frequency in their transmission spectra, where light undergoes diffraction reflection on their periodical structure. The interest in PCs is conditioned both by their interesting physical properties and wide practical applications. As these structures are designed artificially, or in a self organizing manner, they can be prepared with beforehand given properties, which lead to many challenging problems of theoretical and applied character. The optical elements constructed on the basis of PCs result in intelligent, multifunctional tunable optics, which possess many favorable traits, such as their compactness, small losses, high reliability and compatibility with other devices. Cholesteric liquid crystals

(CLCs) are the most representative among the one dimensional (1D) chiral PCs, because they can spontaneously self organize their periodic structure, and their PBG (that exists only for circularly polarized light with the same handedness of the CLC helix); and they can be easily tuned over wide frequency intervals. This polarization-discriminatory filtering characteristic of a CLC is attractive in optical technology. Liquid crystal devices (LCDs) are well-known building blocks of many modern electro-mechanical-magneto-optical systems. Among these special LCDs are: linear polarization rotators [5-7], dynamical wave plate retarders, achromatic [8, 9] pixilated LCD for displays, spatial light modulators, tunable filters [10-12], mirrorless dye lasers [13-15], optical diodes [16-17]. Recently the CLC having various types of defects have been considered from the point of view of generating additional resonance modes in them (see [18-22], and the references cited in [18-22]).

The state of polarization, as the fundamental property of a light wave, has drawn a great deal of attention due to its interesting properties and potential applications. In such polarization sensitive systems as: waveguides [23], coherent detectors [24] and polarization-based switches [25], the state of polarization plays an important role. And in wave-length-division-multiplexing systems, the polarization-mode dispersion, polarization-dependent loss and unpredicted state of polarization drift (due to thermal, mechanical or pressure perturbations) can be accumulated, which present a challenge for long-haul telecommunications [26-29]. To control the state of the optical polarization a number of useful polarization controllers, such as: a squeezed fiber [30], rotating wave-plates [31], electro-optic wave-plates [32], Faraday rotators [33-36], and rotating magnetic field type ones [37] have been reported. An important problem of ellipsometrics (as well as polarimetrics and optoelectronics) is as follows: the elements of optical systems are usually polarization sensitive, and these elements change the polarization state of the light. Each of these elements carries out its own function for a given polarization. Hence, the application of a polarization azimuth stabilizer becomes important. Polarization azimuth stabilizers must satisfy the following condition: the change of the light polarization state at the stabilizer's entrance must not substantially change the polarization state at the stabilizer's exit. The possibilities of polarization azimuth stabilization by an anisotropic/isotropic homogeneous/inhomogeneous plate are discussed in [38-40]. In the present paper, we also investigate the possibilities of using CPCs with a defect as tunable polarization azimuth stabilizers.

Measurements of a light beam polarization state lie on the base of many fundamental experiments of the modern physics, such as the experiments connected with parity violation in atoms [41-44] and detection of magnetic birefringence of vacuum [45], as well as with the applied physics measurement techniques – for instance in magnteometrics [46] and ellipsometrics [47]. The shot noise restricts possibilities of weak changes in the laser beam polarization state. There are also other resources of optical noise, which restrict the polarimetric measurement sensibility. In such cases, the amplification of the light polarization changes – before the light beam had reached the polarimetric detecting system – would lead to an increase of the system's sensibility.

There are many ways of amplification of the polarization azimuth weak changes (see, for instance, [38-41,51-53], as well as the references cited there). In [51], a method is described in which a dichroic plate is used to amplify the polarization plane weak rotations, but the amplification leads to a decrease of the signal intensity. In [52], a way of amplification is offered in which the light beam reflects from the isotropic half space medium. But in this case, too, the intensity change due to the polarization azimuth change is decreased if the amplification coefficient increases. As it is shown in [52], choosing appropriate (larger) beam intensity, one can measure the almost vanishing rotations of the polarization plane, see also [54-118].

Let us note that in certain experiment conditions the signal intensity values are restricted by the linear optics limits, which, in its turn, restricts the possibilities of the above-said methods. Besides, large intensities worsen the polarimetric measurement sensibility, and there are cases when there is no possibility of having large intensities, at all. These make the search of new possibilities of the polarization plane weak rotation amplification very important.

In [39, 40], a simple and effective method is offered for this, namely, the use of an isotropic /anisotropic layer doped with amplifying molecules (luminescent ones, or dye molecules) as a polarization plane rotation amplifier. In [40], some peculiarities of light polarization weak changes are investigated, which take place if light is transmitted through a CLC layer doped with absorbing/or amplifying molecules. It is shown that the CLC layer can be used for amplification of the polarization plane weak rotations.

## II. THE METHOD OF ANALYSIS

The problem is solved by Ambartsumian's layer addition modified method adjusted to solution of such problems (see, [18]). A *CLC layer* with a defect can be treated as a multi-layer system: *CLC(1)-Defect Layer-CLC(2)* (Fig. 1).

The problem of finding the reflected and transmitted wave amplitudes can be presented as follows:

$$\vec{E}_r = \hat{R}\,\vec{E}_i, \quad \vec{E}_t = \hat{T}\,\vec{E}_i, \tag{1}$$

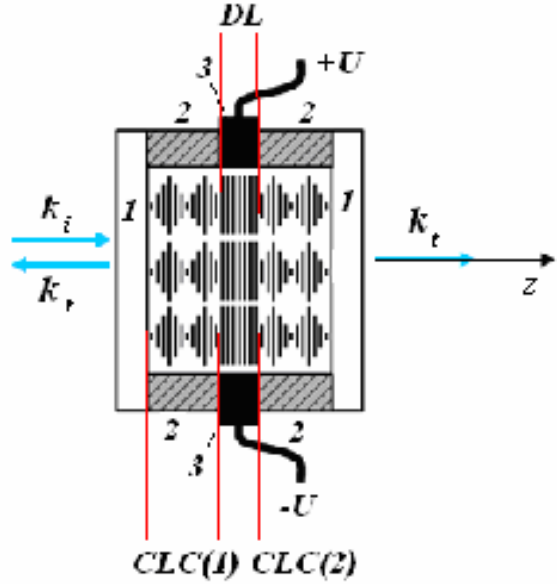

Fig. 1. A sketch diagram of a cell with a chiral liquid crystal with one defect: 1– glass substrates; 2– Teflon fillers; 3–metal electrodes.

where the indices *i*, *r* and *t* denote the incidence, reflected and transmitted fields respectively; $\hat{R}$ and $\hat{T}$ are the reflection and transmission matrices of *CLC(1)-Defect Layer-CLC(2)* system;

$$E_{i,r,t} = E^p_{i,r,t} n_p + E^s_{i,r,t} n_s = \begin{bmatrix} E^p_{i,r,t} \\ E^s_{i,r,t} \end{bmatrix};\ n_p\text{ and } n_s \text{ are orts of } p \text{ and } s \text{ polarizations, respectively.}$$

According to Ambartsumian's layer addition modified method [18], if there is a system consisting of two adjacent (from the left to right) layers, *A* and *B*, then the reflection and transmission matrices of the system, *A+B*, viz. $\hat{R}_{A+B}$ and $\hat{T}_{A+B}$, are determined in terms of similar matrices of its component layers by the matrix equations:

$$\hat{R}_{A+B} = \hat{R}_A + \tilde{\hat{T}}_A \hat{R}_B \left[ \hat{I} - \tilde{\hat{R}}_A \hat{R}_B \right]^{-1} \hat{T}_A,$$
$$\hat{T}_{A+B} = \hat{T}_B \left[ \hat{I} - \tilde{\hat{R}}_A \hat{R}_B \right]^{-1} \hat{T}_A, \qquad (2)$$

where the tilde denotes the corresponding reflection and transmission matrices for the reverse direction of light propagation, and $\hat{I}$ is the unit matrix. The exact reflection and transmission matrices for a finite **CLC layer** (at normal incidence) and a defect (isotropic/anisotropic) layer are well known [42, 43]. First, we attach the defect layer (**DL**) with the **CLC Layer (2)** from the left side, using the matrix Eqs (1). In the second stage, we attach the **CLC Layer (1)** with the obtained **DL-CLC Layer (2)** system.

We pass on to the investigation of possible uses of these systems as polarization azimuth stabilizers with a tunable azimuth.

Let us present the incident light field in the form:

$$E_x = \cos\varphi_i - ie_i \sin\varphi_i \quad \text{and} \quad E_y = \sin\varphi_i + ie_i \cos\varphi_i, \qquad (3)$$

where $\varphi_i$ is the azimuth and $e_i$ is the ellipticity of the incident wave. We investigate the transmitted wave azimuth and ellipticity dependences on the incident wave azimuth and ellipticity for a CLC layer with a defect layer inside. Our calculations show that in the general case the transmitted wave is elliptically polarized and its ellipticity and polarization azimuth essentially change with the incident light azimuth and ellipticity.

For the *x*- and *y*- components of the transmitted wave we have (according to (1)):

$$E_t^x = T_{11} E_x + T_{12} E_y,$$
$$E_t^y = T_{21} E_x + T_{22} E_y \qquad (4)$$

The connection between the azimuths, $\varphi_i$ and $\varphi_t$, of the transmitted waves is defined by the following formula:

$$\text{tg}2\varphi_t = \frac{2\text{Re}\chi}{1-|\chi|^2}, \qquad (5)$$

where $\chi = E_t^y / E_t^x$.

For our case, we have:

$$\chi = \frac{(T_{22} - ie_i T_{21})\text{tg}\varphi_i + (T_{21} + ie_i T_{22})}{(T_{12} - ie_i T_{11})\text{tg}\varphi_i + (T_{11} + ie_i T_{12})}. \qquad (6)$$

Substituting this expression into (5), we get:

$$\varphi_t = \frac{1}{2}\arctg\left(\frac{\alpha_1 \text{tg}^2\varphi_i + \beta_1 \text{tg}\varphi_i + \gamma_1}{\alpha_2 \text{tg}^2\varphi_i + \beta_2 \text{tg}\varphi_i + \gamma_2}\right), \qquad (7)$$

where

$$\alpha_1 = T_{22}T_{12}^* + T_{12}T_{22}^* + e_i^2\left(T_{11}T_{21}^* + T_{21}T_{11}^*\right) + ie_i\left(T_{22}T_{11}^* - T_{11}T_{22}^* + T_{12}T_{21}^* - T_{21}T_{12}^*\right),$$

$$\alpha_2 = |T_{12}|^2 - |T_{22}|^2 - e_i^2\left(|T_{11}|^2 - |T_{21}|^2\right) + ie_i\left(T_{21}T_{22}^* - T_{22}T_{21}^* + T_{12}T_{11}^* - T_{11}T_{12}^*\right),$$

$$\beta_1 = \left(1 - e_i^2\right)\left(T_{22}T_{11}^* + T_{21}T_{12}^* + T_{11}T_{22}^* + T_{12}T_{21}^*\right),$$

$$\beta_2 = \left(1 - e_i^2\right)\left(T_{11}T_{12}^* + T_{12}T_{11}^* - T_{22}T_{21}^* - T_{21}T_{22}^*\right),$$

$$\gamma_1 = T_{21}T_{11}^* + T_{11}T_{21}^* + e_i^2\left(T_{22}T_{12}^* + T_{12}T_{22}^*\right) + ie_i\left(T_{22}T_{11}^* - T_{11}T_{22}^* + T_{12}T_{21}^* - T_{21}T_{12}^*\right), \quad (8)$$

$$\gamma_2 = |T_{11}|^2 - |T_{21}|^2 + e_i^2\left(|T_{12}|^2 - |T_{22}|^2\right) + ie_i\left(T_{21}T_{22}^* - T_{22}T_{21}^* + T_{12}T_{11}^* - T_{11}T_{12}^*\right).$$

The complex conjugate values are denoted by asterisks. The azimuth amplification coefficient, $f$, is the derivative of $\varphi_t$ by $\varphi_i$:

$$f = \frac{d\varphi_t}{d\varphi_i} = \frac{2tg\varphi_i\left(D\alpha_1 - C\alpha_2\right) + \left(D\beta_1 - C\beta_2\right)}{2\cos^2\varphi_i\left(C^2 + D^2\right)}, \quad (9)$$

where

$$C = \alpha_1 tg^2\varphi_i + \beta_1 tg\varphi_i + \gamma_1,$$
$$D = \alpha_2 tg^2\varphi_i + \beta_2 tg\varphi_i + \gamma_2. \quad (10)$$

Another important azimuth amplifying characteristic is the amplifier resolving power, $R$. according to [38], we have for the resolving power:

$$R = \left|\frac{d\varphi_t}{d\varphi_i}\frac{\delta\varphi_i}{\delta\varphi_t}\right| = \sqrt{f^2 \cdot \frac{1-e_t^2}{1+e_t^2} \cdot \frac{1+e_i^2}{1-e_i^2} \cdot \frac{I_t}{I_i}}, \quad (11)$$

where $e_t$ and $e_i$ are the polarization ellipticities of the transmitted/incident waves, respectively $\left(e_t = tg\left(\frac{1}{2}\arcsin\left(\frac{2\,\text{Im}\,\chi}{1+|\chi|^2}\right)\right)\right)$, and $I_t$ and $I_i$ are their intensities.

We investigate the azimuth inhomogeneity peculiarities of the *CLC layer* with an isotropic/anisotropic defect inside. The ordinary and extraordinary refractive indices of the *CLC layer* are taken to be: $n_o = 1.4639$ and $n_e = 1.5133$, the *CLC layer* helix is right handed and its pitch is, $p = 0.42$ μm. These are the parameters of the *CLC cholesteryl-nonanoate–cholesteryl chloride–cholesteryl acetate* (20 : 15 : 6) composition, again, at the temperature $T = 25°$ C. Thus, for the normal light incidence onto a single *CLC layer* – with the right circular polarization – there is a PBG, and the light with the left circular polarization has not any. The ordinary and extraordinary refractive indices of the *defect layer* are taken to be: $n_o^N = 1.4639$ and

$n_e^N = 1.5133$, (i.e. we assume that the defect is caused by an external static electric field, therefore, the defect layer refractive indices coincide with the local indices of the CLC layer), and $n^d = 1.7$ (for the isotropic defect).

### III. THE RESULTS AND THEIR DISCUSSION

As it is known, the polarization plane rotation amplification and polarization azimuth stabilization take place due to the azimuth non-equivalency. This non-equivalency leads to a non-linear relation between $\varphi_t$ and $\varphi_i$. It results an azimuth change $\Delta\varphi_t$ if $\varphi_i$ is changed by $\Delta\varphi_i$ – in certain regions of $\varphi_i$ – and $\Delta\varphi_t \neq \Delta\varphi_i$. There are ranges of $\varphi_i$ where the azimuth amplification coefficient – defined as $f = \dfrac{d\varphi_t}{d\varphi_i}$ – is greater than the unit (here the given device can work as a polarization plane rotation amplifier) and there also are some ranges of $\varphi_i$ where $f \ll 1$ (here the given device can work as a polarization azimuth stabilizer).

As it is shown in [38,39], the period of the $|f|$ with respect $\varphi_i$ is $\pi$ in anisotropic media if the magnetic field and optical activity are absent, moreover, $|f|$ has two identical maxima in the region of $\varphi_i$, [0, $\pi$], which exceed the unit and are symmetrical in respect to the $\varphi_i = \pi/2$ axis.

Besides, the azimuths of $|f|$ maximums coincide with those of $|e_t|$. This explains the fact that the increase of the polarization plane rotation of the amplifier is so small at the maximum amplification.

Polarization plane rotation amplification takes place in the CLC layer, too. But in contrast to the ordinary isotropic/anisotropic media, the number of those parameters on which the amplification coefficient depends is increased in the CLC layer (it gains one parameter more, namely, the helix pitch). Besides, the CLC parameters are easily tuned and therefore the investigation of peculiarities of the azimuthal inhomogeneities in the CLC layer and the CLC layer with a defect inside is very actual. When using a CLC layer, the light transmitted through it has an elliptic polarization, and this ellipticity is usually greater at the azimuths of maximum amplification and it leads to an essential decrease of the resolving power of the azimuth measuring device. As the calculations show, the period of $|f|$ in respect of $\varphi_i$ is $\pi$ in the CLC, too, but if $\varphi_i$ changes from 0 to $\pi$, $|f|$ has only one maximum – and it is greater than the unit. Here, too, the maximum of $|f|$ coincides with the maximum of the azimuth, $|e_t|$.

The calculations also show that of the **CLC** homogeneous layer strongly depend on the parameter α, which characterizes the influence of the dielectric borders: $\alpha = \sqrt{\dfrac{\varepsilon_m}{\varepsilon}}$, where $\varepsilon_m = (\varepsilon_1 + \varepsilon_2)/2$, $\varepsilon_1$ and $\varepsilon_2$ are the **CLC** dielectric local tensor principal values, $\varepsilon$ is the dielectric permittivity of the medium bordering with the **CLC layer** from both sides – this influence is minimum at $\alpha = 1$. Besides, $f$ and $R$ depend on the parameter, $d\delta/p$, where $d$ is the CLC layer thickness, $\delta = (\varepsilon_1 - \varepsilon_2)/(\varepsilon_1 + \varepsilon_2)$, $p$ is the helix pitch (this parameter characterizes diffraction efficiency). As it is shown in [41], the dielectric borders lead to a significant increase of the polarization plane rotation, about 12-25 times, as well as to a likewise increase of the nonreciprocity in the CLC [16]. Then, the local anisotropy influence is maximum at the wavelength equal to the helix pitch, as well as at the Mogen limit (σ>>λ/δ), i.e. if the helix pitch is much more than the wavelength in the medium. In the latter case, the influence of the spirality sharply decreases and the CLC behaves like an ordinary anisotropic medium from the viewpoint of the behavior of $f$.

Again, two maxima (in the function $f$ from $\varphi_i$) appear, if $\varphi_i$ changes from 0 to π, but in contrast to usual anisotropic media and due to weak influence of spirality, these maxima are not symmetrical in respect to the line $\varphi_t = \pi/2$, as they are in anisotropic media. At the other limit when the helix pitch is much less than the wavelength, the anisotropy influence vanishes and the CLC behaves (from the viewpoint of $f$) like an isotropic optically active medium with the permittivity $\varepsilon_m$. In the intermediate region the anisotropy influence oscillates.

As in thick CLC layers in the ideal case ($\alpha = 1$) only the waves with one circular polarization can propagate (the other one with another polarization coinciding with the helix sign is almost completely reflected), the system here does not "remember" the incident light orientation connection with the director orientation, and the amplification almost vanishes, $f \approx 0$, and the thicker the CLC, the more complete is the reflection of the resonance circular reflection and the more is the proximity of $f$ to zero.

But here the mentioned layer cannot work as a polarization azimuth stabilizer, because the ellipticity of the transmitted light equals to the unit (i.e. the transmitted light has circular polarization).

The difference of α from the unit makes the transmitted light elliptically polarized and its polarization ellipticity in the PBG depends weakly on the polarization azimuth angle of the incident light (due to multiple reflection), and $f$ becomes a little different from zero. If the CLC thickness increases, $f$ decreases in the PBG. At the borders of the PBG, when both circular polarizations propagate in the CLC, and the anisotropy influence is great, the amplification is maximum. Away from the PBG borders, this maximum amplification decreases with oscillations, but it is always more than the unit. If $d\delta/p<1$, the anisotropy influence sharply decreases, therefore, $f \approx 1$, in this case. The amplification is also absent at $d\Delta n\lambda<<1$, in ordinary anisotropy media.

In the intermediate case, $d\delta/p\sim1$, the waves of both circular polarization propagate in the PBG (and the reflectance coefficient of the diffracting polarization in the PBG is of the order of 0.5). It leads to the maximum amplification at the PBG center.

Let us pass on to investigation of the amplification peculiarities in the CLC layer with a defect inside. When there is a thin defect in the CLC structure, it leads to emergence of a defect mode in the PBG. It maintains itself in the form of a dip in the reflection spectrum of the light with the right circular polarization (diffracting polarization), and in the form of a peak in the reflection spectrum of the light with the left circular polarization (non-diffracting polarization). The CLC helix is a right one.

As our calculations show, in this case the new azimuth inhomogeneity peculiarities emerge at the defect mode wavelength and nearby this wavelength. In Fig. 2 the following dependences are presented: the amplification coefficient, $f$, dependence (curve 1); the polarization ellipticity, $e_t$, dependence (curve 2); the resolving power, $R$, dependence (curve 3); the relative intensity, $I_t/I_i$, dependence (curve 4) on the incident light polarization azimuth, $\varphi_i$, for various thicknesses: of the defect layer; of the CLC layer, and for various defect layer locations.

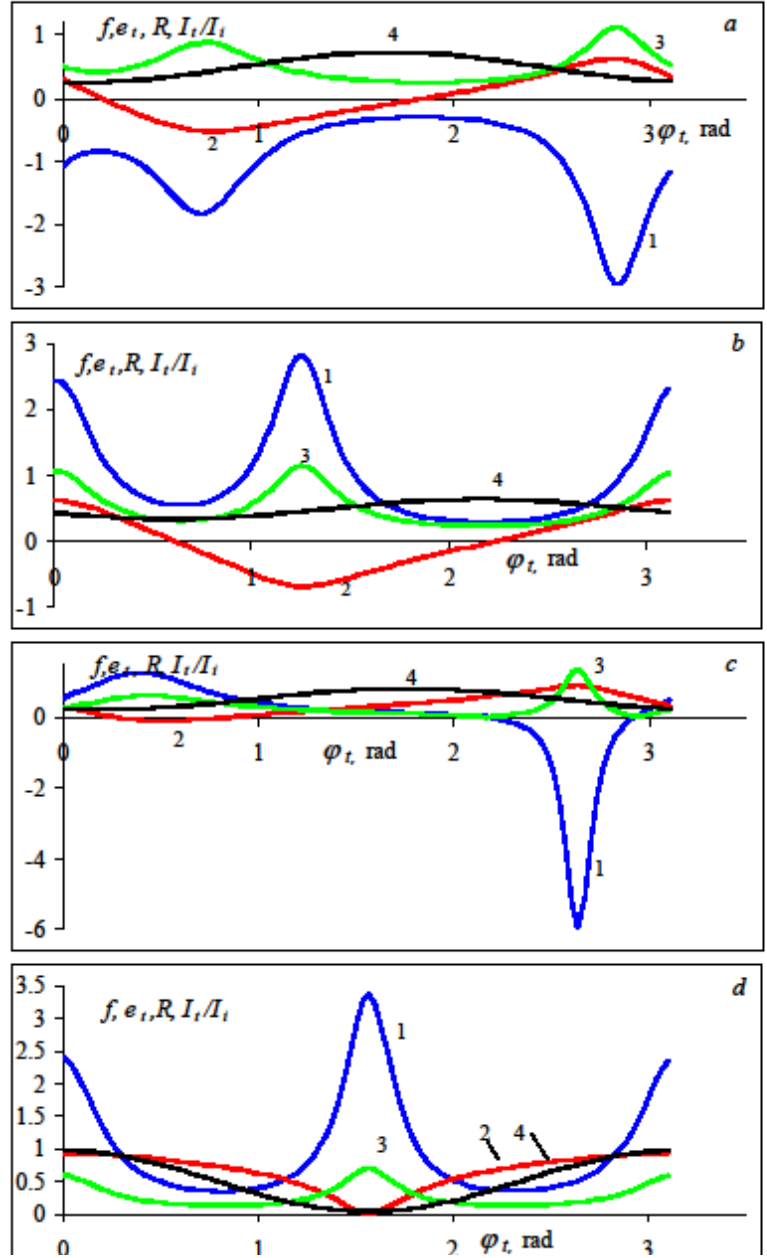

Fig. 2. The dependences of: the amplification coefficient, $f$ (curve 1); of the polarization ellipticity, $e_t$, (curve 2); of the resolving power, $R$, (curve 3); and of the relative intensity, $I_t/I_i$ (curve 4) on the incident light polarization azimuth, $\varphi_i$, for the defect layer and **CLC layer** various thicknesses and for various defect layer locations, for the following cases: **a.** The defect is anisotropic and is at **CLC center**, and: the defect layer thickness is $d_d=1.86$ μm, the incident light wavelength is $\lambda=0.62$ μm. **b.** the defect is isotropic and: is at **CLC layer** center, $d_d=1.86$ μm, $\lambda=0.6175$ μm. **c.** The defect is isotropic and: is at CLC layer center, $d_d=0.2$ μm, $\lambda=0.6162$ μm. **d.** the defect is anisotropic (and it is a half-wave plate) and it is at the CLC layer right edge, $d_d=2.5$ μm, $\lambda=0.625$ μm.

In Fig 2a, the defect is anisotropic and is located at the CLC layer center. In Fig. 2b and 2c, the defect is isotropic and is located at the CLC layer center. In Fig. 2d the defect is anisotropic (and it is a half-wave plate) and it is located at the CLC layer right edge.

The calculations were carried out for the defect mode wavelength. As it is seen from the figures, in this case – in contrast to the homogeneous CLC layer case – again, two maxima of the $f$ on $\varphi_i$ dependence emerge, if $\varphi_i$ varies from 0 to π, as they do in ordinary anisotropic media. But in contrast to the latter, these maxima are not symmetrical in respect to the line, $\varphi_t = \pi/2$.

The azimuths of the maxima of $|e_t|$ mainly coincide with those of $|f|$. But in contrast to the homogeneous CLC layer, or the inhomogeneous plate, this is not a rule (see Fig. 2c and 2d). In this case, a situation is possible, when $|e_t|$ has a minimum with $|e_t| \approx 0$ at the maximum amplification azimuth. But here not only the great values of the solving power are at the maximum amplification azimuth, but the transmitted light intensity is minimum. Such situation takes place when light reflects from an anisotropic plate at the Brewster angle [52], as well as in the case of light transmittance through a dichroic plate [51].

Now we pass on to the investigation of the polarization azimuth stabilization.

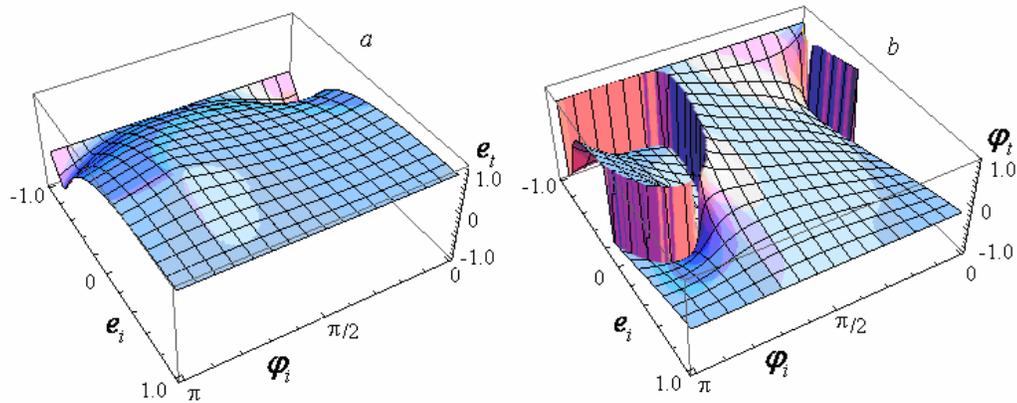

Fig. 3. The 3D plots of the transmitted wave ellipticity, $e_t$, (*a*) and the azimuth, $\varphi_t$, (*b*) on the incident wave ellipticity, $e_i$, and the azimuth, $\varphi_i$. The defect is anisotropic with the thickness: $d^d = 0.1 \mu m$, $\alpha = 1$. The incident light wavelength is: λ = 0.625 μm.

In Fig. 3, the dependences of the transmitted wave ellipticity, $e_t$, (a), and the azimuth, $\varphi_t$, (b), on the ellipticity, $e_i$, and the azimuth, $\varphi_i$, of the incident wave are presented. As can be seen from the figure, in this case, indeed, the ellipticity and polarization azimuth of the transmitted wave vary significantly when the incident light azimuth and ellipticity change, and practically no azimuth stabilization takes place. As our calculations show, we have good results when the defect layer is a quarter-wave plate and is located at the **CLC layer** left edge, i.e. if light is transmitted through the **NLC-CLC system.** Indeed, as can be seen from the figure, in this case, the transmitted wave ellipticity practically does not depend on the incident light azimuth and ellipticity. Some change of the transmitted wave polarization azimuth is observed nearby the point, $e_i = 0$, at $\varphi_i = \pi/2$ (see Fig. 4).

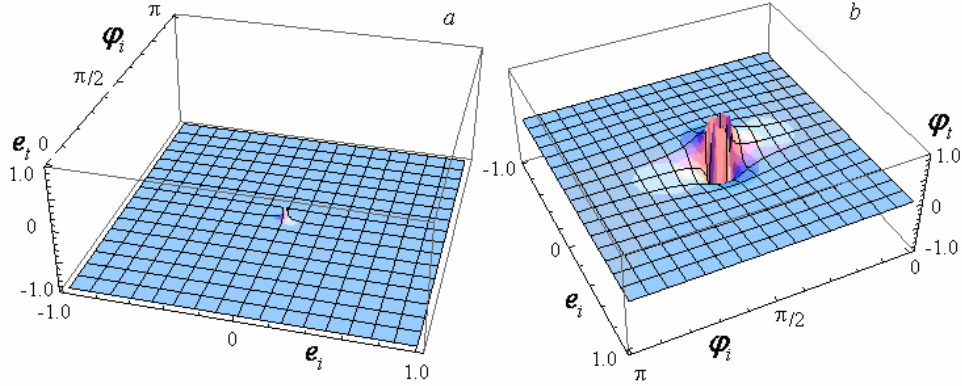

Fig. 4. The 3D plots of the transmitted wave ellipticity, $e_t$, (*a*) and the azimuth, $\varphi_t$, (*b*) on the incident wave ellipticity, $e_i$, and the azimuth, $\varphi_i$, for the **NLC-CLC system. NLC layer** is a quarter-wave plate. $\alpha = 1$. The incident light wavelength is $\lambda = 0.625$ μm.

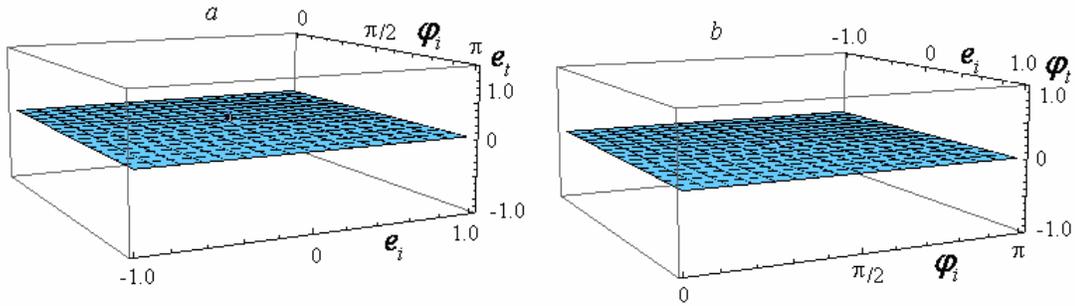

Fig. 5. The 3D plots of the transmitted wave ellipticity, $e_t$, (*a*) and the azimuth, $\varphi_t$, (*b*) on the incident wave ellipticity, $e_i$, and the azimuth, $\varphi_i$, for the **NLC-CLC-NLC system. NLC layers** are quarter-wave plates. $\alpha = 1$. The incident light wavelength is $\lambda = 0.625$ μm.

Detailed investigations show that better results are obtained for the system **NLC-CLC-NLC,** for the case if the **NLC layers** are quarter-wave plates. In Fig. 5, the same dependences as in Fig. 3 are presented for this case. As it is seen from the figure, the system can work as an ideal polarization azimuth stabilizer in this case. The transmitted wave azimuth does not vary if the incident wave azimuth and ellipticity change. Moreover, the transmitted wave polarization azimuth can be changed through the change of the angle, $\varphi_t$, between the **NLC layer** optical axis and the incident light polarization azimuth. Since this angle can be changed mechanically (for instance, if the anisotropic layer is solid) or by an external static electrical field, we have an ideal polarization azimuth stabilizer with a tunable azimuth of stabilization. This system can work as a polarization converter. We can continuously rotate the azimuth of the linearly polarized light by continuous change of the **NLC layer** optical axes orientation. This system could be used as a polarization axis finder, or phase modulator for analyzing biological tissues and polarizing materials, or in diffractive optics, and in other optical elements.

## 4. Conclusions

We investigated the peculiarities of polarization plane rotation amplification when light transmits through a **CLC layer** with a defect inside. We showed that the system gains new azimuth inhomogeneity peculiarities at the defect mode, which distinguishes the system from the homogeneous **CLC layer**. In particular, we showed that for certain parameters of the problem, the polarization azimuth maximum amplification can coincide with the polarization azimuth of the ellipticity module minimum. But as in the cases described in [51,52], in this case, an improvement of the solving power does not happen. Probably, from this side of view, the amplification method described in [39] can be more perspective.

We also investigated the peculiarities of the polarization azimuth stabilization by the CLC system with a defect inside, and we offer a tunable polarization azimuth stabilizer.

## References

1. J. Joannopoulos, R. Meade, J. Winn. Photonic Crystals, Princeton: Princeton Univ. (1995).
2. K. Sakoda, Optical Properties of Photonic Crystals, Berlin: Springer (2001).
3. S. G. Johnson, J. Joannopoulos. Photonic Crystals: The Road from Theory to Practice, Boston: Kluwer (2002).
4. Soukoulis, C. M. (Ed.). Photonic Crystals and Light Localization in the 21st Century. NATO Science Series C: Vol. 563. 2001, 616 p.
5. C. Ye. Opt. Eng. **34**, 3031-3035. (1995).
6. F. Yang, L. Ruan, S. A. Jewell, and J. R. Sambles. Opt. Express. **15**, 4192-4197 (2007).

7. I. Abdulhalim, and A. Safrani. Opt. Lett. **34**, 1801-1803 (2009).

8. M. D. Lavrentovich, T. A. Sergan, and J. R. Kelly. Opt. Lett. **29**, 1411-1413 (2004).

9. Q-H. Wang, T. X. Wu, X. Zhu, and S-T. Wu. Liquid Crystals **31**, 535-539 (2004).

10. D-K. Yang, and S-T. Wu, Fundamentals of Liquid Crystals .Wiley (2006).

11. A. Lakhtakia, and M. McCall. Opt. Commun. **168**, 457-465 (1999).

12. H. Sarkissian, B. Ya. Zeldovich, and N. V. Tabiryan, Opt. Lett. **31**, 1678-1680 (2006).

13. V. I. Kopp, B. Fan, H. K. M. Vithana, and A. Z. Genack. Opt. Lett. **23**, 1707-1709 (1998).

14. A. F. Munoz, P. Palffy-Muhoray, and B. Taheri. Opt. Lett., **26**, 804-806 (2001).

15. T. Matsui, R. Ozaki, K. Funamoto, M. Ozaki, and K. Yoshino. Appl. Phys. Lett. **81**, 3741-3743 (2002).

16. A. H. Gevorgyan. Tech. Phys. **47**, 1008-1013 (2002).

17. M. H. Song , N. Y. Ha, K. Amemiya, B. Park, Y. Takanishi, K. Ishikawa, J. W. Wu, S. Nishimura, T. Toyooka, and H. Takezoe. Adv. Mater. **18**, 779-783 (2006).

18. A. H. Gevorgyan, M. Z. Harutyunyan. *Phys. Rev. E.* **76**, 031701-9 (2007).

19. A. H. Gevorgyan, K. B. Oganesyan, E. M. Harutyunyan, S. O. Arutyunyan. *Opt. Communn*. **283**, 3707-3713 (2010).

20. A. H. Gevorgyan. *Opt. Commun.*, **281**, 5097-5103 (2008).\

21. A. H. Gevorgyan, and M. Z. Harutyunyan. *J. Mod. Opt.,* **56**, 1163-1173(2009).

22. A. H. Gevorgyan. Phys. Rev. E., **83**, 011702 (2011).

23. R. Alferness. IEEE J. Quantum Electron. **17,** 965-969 (1981).

24. N. G. Walker and G. R. Walker. J. Lightwave Technol. **8** (1990) 438-458.

25. A. V. Krishnamoorthy, F. Xu, J. E. Ford, and Y, Fainman, Appl. Opt. **36** (1997) 997-1010.

26. C. D. Poole, Opt. Lett. **14**  (1989) 523-525.

27. T. Ono, S. Yamazaki, H. Shimizu, and K. Emura, J. Lightwave Technol. **12** (1994) 891-898.

28. R. Khosravani, S. A. Havstad, Y. W. Song, P. Ebrahimi, and A. E. Willner, IEEE. Photon Technol. Lett. **13** (2001) 1370-1372.

29. R. Noe, D. Sandel, M. Yoshida-Dierolf, S. Hinz, V. Mirvoda, A. Schopflin, C Gungener, E. Gottwald, C. Scheerer, G. Fischer, T. Weyrauch, and W. Haase, J. Lightwave Technol. **17** (1999) 1602-1616.

30. W. H. J. Aarts and G. –D. Khoe, J. Lightwave Technol. **7** (1989) 1033-1043.

31. F. Heismann, J. Lightwave Technol. **12** (1994) 690-699.

32. H. Shimizu and K. Kaede, Electron. Lett. **24** (1988) 412-413.

33. D. Goldberg, Z. Zalevsky, G. Shabtay, D. Abraham, and D. Mendlovich, J. Opt. A., **6** (2004) 98-105.

34. X. S. Yao, L. Yan, and Y. Shi, Opt. Lett. **30** (2005) 1324-1326.

35. Y. Zhang, C. Yang, S. Li, H. Yan, J. Yin, C. Gu, and G. Jin, Opt. Express. **14** (2006) 3484-3490.

36. L. Chen, and W. She, Opt. Express. **15** (2007) 15589-155994.

37. T. Saitoh and S. Kinugawa, Photon. Technol. Lett. **15** (2003) 1404-1406.

38. G. A. Vardanyan, A.H. Gevorgyan, J. Contemp. Phys. (Acad. Sci. Arm.). **31** (1996) 34-42.

39. A. H.Gevorgyan, A. Grigoryan, A. Kocharian, A.Zh. Khachatryan, L. O. Mikaelyan, A. M. Sedrakyan, G. A. Vardanyan. Optik, **117,** No 7, pp. 309-316 (2006).

40. A. A. Gevorgyan, A. M. Sedrakyan, and A. Zh. Khachatryan. Journal of Optical Technology, **75,** No 2, pp. 69-74 (2008).

41. A. H. Gevorgyan, M. Z. Harutyunyan, S. A. Mkhitaryan, E. A. Santrosyan and G. A. Vardanyan. Optik. **121**, 39-44 (2010).

42. A. H. Gevorgyan. *Opt. Spectrosc*., **89**, 631-638 (2000).

43. H. Wohler, M. Fritsch, G. Hass, D. A. Mlynski. *J. Opt. Soc. Am. A*, **8**, 536-540 (1991).

44. M.-A. Bouchiat, C. Bouchia, Rep. Prog. Phys. **60** (1997) 1351–1397.

45. M.J.D. Macpherson, K.P. Zetie, R.B. Warrington, D.N. Stacey, J.P. Hoare, Phys. Rev. Lett. **67** (1991) 2784 - 2787.

46. D. M. Meekhof, P. Vetter, P. K. Majumder, S. K. Lamoreaux, E. N. Fortson, Phys. Rev. Lett. **71** (1993) 3442-3445.

47. N. H. Edwards, S. J. Phipp, P. E. G. Baird, S Nakayama, Phys. Rev. Lett. **74** (1995) 2654-2657.

48. D. Bakalov, F Brandi, G. Cantatore, G. Carugno, S. Carusotto, F. Della Valle, A. M. De Riva, U. Gastaldi, E. Iacopini, P. Micossi, E. Milotti, R. Onofrio, R. Pengo, F.

Perrone, G. Petrucci, E. Polacco, C. Rizzo, G. Ruoso, E. Zavattini and G. Zavattini, Quantum Semiclassic. Opt. **10** (1998) 239-250.

49. I. K. Kominis, T.W.Kornack, J. C. Allred, M. V. Romalis, Nature (London) **422** (2003) 596-599.

50. R. M. A. Azzam, N. M. Bashara, Ellipsometry and polarized light, North-Holland, New York, 1977.

51. V. S. Zapasskii, Journal of Applied Spectroscopy **37** (1982) 181-196.

52. K. K. Svitashev, G. Khasanov, Opt. Spectrosc. **54** 5(1982) 38-539.

53. M. Lintz, J. Guena, M.-A. Bouchiat, D. Chauvat, Rev. Sci. Inst. **76** (2005) 043102

54. Fedorov, M.V., Oganesyan, K.B., Prokhorov, A.M., Appl. Phys. Lett., **53**, 353 (1988).

55. Oganesyan K.B., Prokhorov A.M., Fedorov M.V., Sov. Phys. JETP, **68,** 1342 (1988).

56. Petrosyan M.L., Gabrielyan L.A., Nazaryan Yu.R., Tovmasyan G.Kh., Oganesyan K.B., Laser Physics, **17**, 1077 (2007).

57. E.A. Nersesov, K.B. Oganesyan, M.V. Fedorov, Zhurnal Tekhnicheskoi Fiziki, **56**, 2402 (1986).

58. EM Sarkisyan, KG Petrosyan, KB Oganesyan, VA Saakyan, NSh Izmailyan, and CK Hu, Laser Physics, **18**, 621 (2008).

59. A.H. Gevorkyan, K.B. Oganesyan, E.M. Arutyunyan. S.O. Arutyunyan, Opt. Commun., **283**, 3707 (2010).

60. D.N. Klochkov, A.I. Artemiev, K.B.Oganesyan, Y.V.Rostovtsev, M.O.Scully, C.K. Hu. Physica Scripta, **T 140**, 014049 (2010).

61. K.B. Oganesyan, J. Contemp. Phys., **50,** 123 (2015).

62. K.B. Oganesyan, J. Contemp. Phys., **50,** 312 (2015).

63. D.N. Klochkov, A.I. Artemiev, K.B. Oganesyan, Y.V.Rostovtsev, C.K. Hu. J. Modern Optics, **57**, 2060 (2010).

64. Zaretsky, D.F., Nersesov, E.A., Oganesyan, K.B., and Fedorov, M.V., Sov. J. Quantum Electronics, **16**, 448 (1986).

65. A.H. Gevorgyan**,** M.Z. Harutyunyan, G.K. Matinyan, K.B. Oganesyan, Yu.V. Rostovtsev, G. Kurizki and M.O. Scully**,** Laser Physics Lett., **13,** 046002 (2016).

66. G.A. Amatuni, A.S. Gevorkyan, S.G. Gevorkian, A.A. Hakobyan, K.B. Oganesyan, V. A. Saakyan, and E.M. Sarkisyan, Laser Physics, **18** 608 (2008).

67. K.B. Oganesyan. Laser Physics Letters, **12**, 116002 (2015).

68. A.I. Artemyev, M.V. Fedorov, A.S. Gevorkyan, N.Sh. Izmailyan, R.V. Karapetyan, A.A. Akopyan, K.B. Oganesyan, Yu.V. Rostovtsev, M.O. Scully, G. Kuritzki, J. Mod. Optics, **56**, 2148 (2009).

69. A.H. Gevorgyan, K.B.Oganesyan, M.Z..Harutyunyan, M.S.Rafaelyan, Optik, **123,** 2076 (2012)**.**

70. A.H. Gevorgyan, K.B. Oganesyan, G.A.Vardanyan, G. K. Matinyan, Laser Physics, **24**, 115801 (2014).

71. A.S. Gevorkyan, K.B. Oganesyan, Y.V. Rostovtsev, G. Kurizki, Laser Physics Lett., **12**, 076002, (2015).

72. DN Klochkov, AH Gevorgyan, NSh Izmailian, KB Oganesyan, J. Contemp. Phys., **51,** 237 (2016).

73. K.B. Oganesyan, M.L. Petrosyan, M.V. Fedorov, A.I. Artemiev, Y.V. Rostovtsev, M.O. Scully, G. Kurizki, C.-K. Hu, Physica Scripta, **T140**, 014058 (2010).

74. Oganesyan, K.B., Prokhorov, A.M., and Fedorov, M.V., ZhETF, **94**, 80 (1988); Oganesyan K B, Prokhorov A M and Fedorov M V Zh. Eksp. Teor. Fiz., **53,** 80 (1988).

75. K.B. Oganesyan, J. of Contemporary Physics, **51**, 307 (2016).

76. A.H. Gevorgyan , K.B. Oganesyan, Optics and Spectroscopy, **110**, 952 (2011).

77. K.B. Oganesyan, J. Mod. Optics, **62,** 933 (2015).

78. K.B. Oganesyan. Laser Physics Letters, **13**, 056001 (2016).

79. A.H. Gevorgyan**,** M.Z. Harutyunyan, G.K. Matinyan, K B Oganesyan, Yu.V. Rostovtsev, G. Kurizki and M.O. Scully**,** Laser Physics Lett., **13,** 046002 (2016).

80. K.B. Oganesyan, J. Mod. Optics, **61,** 763 (2014).

81. K.B. Oganesyan, Nucl. Instrum. Methods A **812,** 33 (2016).

82. V.V. Arutyunyan, N. Sh. Izmailyan, K.B. Oganesyan, K.G. Petrosyan and Cin-Kun Hu, Laser Physics, **17**, 1073 (2007).

83. A.H. Gevorgyan, K.B.Oganesyan, E.M.Harutyunyan, S.O.Harutyunyan, Modern Phys. Lett. B, **25**, 1511 (2011).

84. A.H. Gevorgyan**,** K.B. Oganesyan, Laser Physics Lett., **12,** 125805 (2015).

85. K.B. Oganesyan, J. Contemp. Phys., **51,** 10 (2016).

86. D.N. Klochkov, K.B. Oganesyan, E.A. Ayryan, N.Sh. Izmailian, J. of Modern Optics, **63,** 653 (2016).

87. A.H. Gevorgyan, K.B. Oganesyan, J. of Contemporary Physics, **45,** 209 (2010)**.**

88. A.S. Gevorkyan, A.A. Grvorgyan, K.B. Oganesyan, G.O. Sargsyan, N.V. Saakyan, Physica Scripta, **T140**, 014045 (2010).

89. A.S. Gevorkyan, A.A. Gevorkyan, K.B. Oganesyan, Phys. Atom. Nuclei, **73,** 320 (2010).

90. D.N. Klochkov, K.B. Oganesyan, Y.V. Rostovtsev, G. Kurizki, Laser Physics Lett., **11**, 125001 (2014).

91. Zh.S. Gevorkian, K.B. Oganesyan, Laser Physics Lett., **13**, 116002, (2016).

92. K.B. Oganesyan, J. Mod. Optics, **61,** 1398 (2014).

93. D. N. Klochkov, A. I. Artemyev, K. B. Oganesyan, Y. V. Rostovtsev, M. O. Scully, Chin-Kun Hu, Journal of Physics: Conference Series **236,** 012022 (2010).

94. K.B. Oganesyan, A.H. Gevorgyan, G.A. Vardanyan, R.V. Karapetyan, Proceedings of SPIE, 9182-44 (2014).

95. K.B. Oganesyan, **A**.H. Gevorgyan, G.A. Vardanyan, R.V. Karapetyan, Proceedings of SPIE, 9182-47 (2014).

96. E.M. Sarkisyan, Zh.S. Gevorkian, K.B. Oganesyan, V.V. Harutyunyan, V.A. Saakyan, S. G. Gevorgyan, J.Verhoeven, M.V. Fedorov, A.I. Artemiev, S.M. Fedorov, Laser Physics **17**, 1080 (2007).

97. E.A. Ayryan, A.H. Gevorgyan, K.B. Oganesyan, arXiv:1611.04094 (2016).

98. E.A. Ayryan, A.H. Gevorgyan, N.Sh. Izmailian, K.B. Oganesyan, arXiv:1611.06515 (2016).

99. K.B. Oganesyan, arXiv:1611.08774 (2016).

100. I.V. Dovgan, K.B. Oganesyan, arXiv:1612.04608 (2016).

101. L.A.Gabrielyan, Y.A.Garibyan, Y.R.Nazaryan, K.B. Oganesyan, M.A.Oganesyan, M.L.Petrosyan, E.A. Ayryan, arXiv:1701.00916 (2017).

102. E.A. Ayryan, M. Hnatic, K.G. Petrosyan, A.H. Gevorgyan, N.Sh. Izmailian, K.B. Oganesyan, arXiv: 1701.07637 (2017).

103. E.A. Ayryan, K.G. Petrosyan, A.H. Gevorgyan, N.Sh. Izmailian, K.B. Oganesyan, arXiv: 1702.03209 (2017).

104. E.A. Ayryan, K.G. Petrosyan, A.H. Gevorgyan, N.Sh. Izmailian, K.B. Oganesyan, arXiv: 1703.00813 (2017).

105. A.H. Gevorgyan, K.B. Oganesyan, E.A. Ayryan, Yu.V. Rostovtsev. Arxiv:1703.03570 (2017).

106. A.H. Gevorgyan, K.B. Oganesyan, E.A. Ayryan, Yu.V. Rostovtsev. Arxiv:1703.07637 (2017).

107. A.H. Gevorgyan, K.B. Oganesyan, E.A. Ayryan, M. Hnatic, J.Busa, E. Aliyev, A.M. Khvedelidze, Yu.V. Rostovtsev, G. Kurizki, arXiv:1703.03715 (2017).

108. A.H. Gevorgyan, K.B. Oganesyan, E.A. Ayryan, Michal Hnatic, Yuri V. Rostovtsev, arXiv:1704.01499 (2017).

109. E.M. Sarkisyan, A.A. Akopyan, K.B. Oganesyan, K.G. Petrosyan, V.A. Saakyan, Laser Physics, **19,** 881 (2009).

110. K.B. Oganesyan, M.L. Petrosyan, YerPHI-475(18) – 81, Yerevan, (1981).

111. M.V. Fedorov, K.B. Oganesyan, IEEE J. Quant. Electr, vol. **QE-21**, p. 1059 (1985).

112. D.F. Zaretsky, E.A. Nersesov, K.B. Oganesyan, M.V. Fedorov, Kvantovaya Elektron. **13** 685 (1986).

113. A.H. Gevorgyan, K.B. Oganesyan, R.V. Karapetyan, M.S. Rafaelyan, Laser Physics Letters, **10**, 125802 (2013).

114. M.V. Fedorov, E.A. Nersesov, K.B. Oganesyan, Sov. Phys. JTP, **31,** 1437 (1986).

115. K.B. Oganesyan, M.V. Fedorov, *Zhurnal Tekhnicheskoi Fiziki*, **57**, 2105 (1987).

116. A.H. Gevorgyan, K.B. Oganesyan, R.V. Karapetyan, M.S. Rafaelyan, Laser Physics Letters, **10**, 125802 (2013).

117. M.L. Petrosyan, L.A. Gabrielyan, Yu.R. Nazaryan, G.Kh. Tovmasyan, K.B. Oganesyan, J. Contemp. Phys., **42**, 38 (2007).

118. D.N. Klochkov, K.B. Oganesyan**,** E.A. Ayryan, N.Sh. Izmailian, J. of Modern Optics, **63,** 653(2016).